\documentclass[journal]{IEEEtran}
\usepackage{lineno}
\usepackage{eurosym}

\usepackage{color,soul}

\usepackage{xcolor}

\ifCLASSINFOpdf
  \usepackage[pdftex]{graphicx}
	\usepackage{epstopdf}
  \DeclareGraphicsExtensions{.pdf,.jpeg,.png}
\else
\fi
\usepackage{booktabs}
\usepackage{colortbl}
\usepackage{amsbsy}

\usepackage{amssymb}

\usepackage{url}

\begin{document}
%
\title{Condition monitoring and early diagnostics methodologies for hydropower plants}
%
%
%

\author{Alessandro~Betti, 
Emanuele~Crisostomi,~\IEEEmembership{Senior Member,~IEEE,}
Gianluca~Paolinelli,
Antonio~Piazzi,
Fabrizio~Ruffini,
 and~Mauro~Tucci,~\IEEEmembership{Senior Member,~IEEE}
\thanks{A. Betti, A. Piazzi and F. Ruffini are with i-EM S.r.l, Via Lampredi 45, 57121 Livorno, Italy.}
\thanks{E. Crisostomi and M. Tucci are with the Department of Energy, Systems, Territory and Constructions Engineering, University of Pisa, Pisa, Italy. Email: emanuele.crisostomi@unipi.it.}
\thanks{G. Paolinelli is with Pure Power Control s.r.l., Via Carbonia, 2, 56021, Navacchio, Pisa, Italy.}
}

\maketitle

\begin{abstract}
Hydropower plants are one of the most convenient option for power generation, as they generate energy exploiting a renewable source, they have relatively low operating and maintenance costs, and they may be used to provide ancillary services, exploiting the large reservoirs of available water. The recent advances in Information and Communication Technologies (ICT) and in machine learning methodologies are seen as fundamental enablers to upgrade and modernize the current operation of most hydropower plants, in terms of condition monitoring, early diagnostics and eventually predictive maintenance. While very few works, or running technologies, have been documented so far for the hydro case, in this paper we propose a novel Key Performance Indicator (KPI) that we have recently developed and tested on operating hydropower plants. In particular, we show that after more than one year of operation it has been able to identify several faults, and to support the operation and maintenance tasks of plant operators. Also, we show that the proposed KPI outperforms conventional multivariable process control charts, like the Hotelling $t_2$ index. 
\end{abstract}


%
\IEEEpeerreviewmaketitle

\section{Introduction}

\subsection{Motivation}
\IEEEPARstart{A}{s} power generation from renewable sources is increasingly seen as a fundamental component in a joint effort to support decarbonization strategies, hydroelectric power generation is experiencing a new golden age. In fact, hydropower has a number of advantages compared to other types of power generation from renewable sources. Most notably, 
hydropower generation can be ramped up and down, which provides a valuable source of flexibility for the power grid, for instance, to support the integration of power generation from other renewable energy sources, like wind and solar. In addition, water in hydropower plants' large reservoirs may be seen as an energy storage resource in low-demand periods and transformed into electricity when needed \cite{Helseth2016,Hjelmeland2019}. Finally, for large turbine-generator units, the mechanical-to-electrical energy conversion process can have a combined efficiency of over 90\% \cite{Bolduc2014}. Accordingly, in 2016, around 13\% of the world's consumed electricity was generated from hydropower\footnote{\url{https://www.iea.org/statistics/balances/}}. In addition, hydropower plants have provided for more than 95\% of energy storage for all active tracked storage installations worldwide\footnote{\url{http://www.energystorageexchange.org/projects/data_visualization}}.

In addition to the aforementioned advantages, hydroelectric power plants are also typically characterized by a long lifespan and relatively low operation and maintenance costs, usually around 2.5\% of the overall cost of the plant. However, according to \cite{IHA_db}, by 2030 over half of the world's hydropower plants will be due for upgrade and modernization, or will have already been renovated. Still according to \cite{IHA_db}, the main reason why major works seem around the corner is that most industries in this field wish to adopt best practices in operations and asset management plans, or in other words, share a desire for optimized performance and increased efficiency. In combination with the quick pace of technological innovation in hydropower operations and maintenance, together with the increased ability to handle and manage big amounts of data, a technological revolution of most hydropower plants is expected to take place soon. 

\subsection{State of the art}
In hydropower plants, planned periodic maintenance has been for a long time the main, if not the only one, adopted maintenance method. Condition monitoring procedures have been often reserved for protection systems, leading to shutting down the plants when single monitored signals exceeded pre-defined thresholds (e.g., bearings with temperature and vibration protection). In this context, one of the earliest works towards predictive maintenance has been \cite{Jiang2008}, where Artificial Neural Networks (ANNs) were used to monitor, identify and diagnose the dynamic performance of a prototype of system. Predictive maintenance methods obviously require the measurement and storage of all the relevant data regarding the power plant. An example of early digitalization is provided in \cite{Li2009} where a Wide Area Network for condition monitoring and diagnosis of hydro power stations and substations of the Gezhouba Hydro Power Plant (in China) was established. Thanks to measured data, available in real-time, more advanced methods that combine past history and domain knowledge can provide more efficient monitoring services, advanced fault prognosis, short- and long-term prediction of incipient faults, prescriptive maintenance tools, and residual lifetime and future risk assessment. Benefits of this include, among other things, preventing (possibly severe) faults from occurring, avoiding unnecessary replacements of components, more efficient criteria for scheduled maintenance.

The equipment required for predictive maintenance in hydro generators is also described in \cite{Ribeiro2014}, where the focus was to gain the ability to detect and classify faults regarding the electrical and the mechanical defects of the generator-turbine set, through a frequency spectrum analysis. More recent works (e.g., \cite{Selak2014}) describe condition monitoring and fault-diagnostics (CMFD) software systems that use recent Artificial Intelligence (AI) techniques (in this case a Support Vector Machine (SVM) classifier) for fault diagnostics. In \cite{Selak2014} a CMFD has been implemented on a hydropower plant with three Kaplan units. Another expert system has been developed for an 8-MW bulb turbine downstream irrigation hydropower plant in India, as described in \cite{Buaphan2017}.
An online temperature monitoring system was developed in \cite{Milic2013}, and an artificial neural network based predictive maintenance system was proposed in \cite{Chuang2004}. The accuracy of early fault detection system is an important feature for accurate reliability modeling \cite{Khalilzadeh2014}.

\subsection{Paper contribution}
In this paper we propose a novel Key Performance Indicator (KPI) based on an appropriately trained Self-Organizing Map (SOM) for condition monitoring in a hydropower plant. In addition to detecting faulty operating conditions, the proposed indicator also identifies the component of the plant that most likely gave rise to the faulty behaviour. Very few works, as from the previous section, address the same problem, although there is a general consensus that this could soon become a very active area of research \cite{IHA_db}. In this paper we show that the proposed KPI performs better than a standard multivariate process control tool like the Hotelling control chart, over a test period of more than one year (from April 2018 to July 2019).

This paper is organized as follows: Section \ref{case_study} describes more in detail the case study of interest, and the data used to tune the proposed indicator. Section \ref{Methodologies} illustrates the proposed indicator. Also, the basic theory of the Hotelling multivariate control chart is recalled, as it is used for comparison purposes. The obtained results are provided and discussed in Section \ref{Results}. Finally, in Section \ref{Conclusion} we conclude our paper and outline our current lines of research in this topic.

\section{Case study}
\label{case_study}
\begin{figure}[ht]
	\centering
	\includegraphics[width=0.3\textwidth]{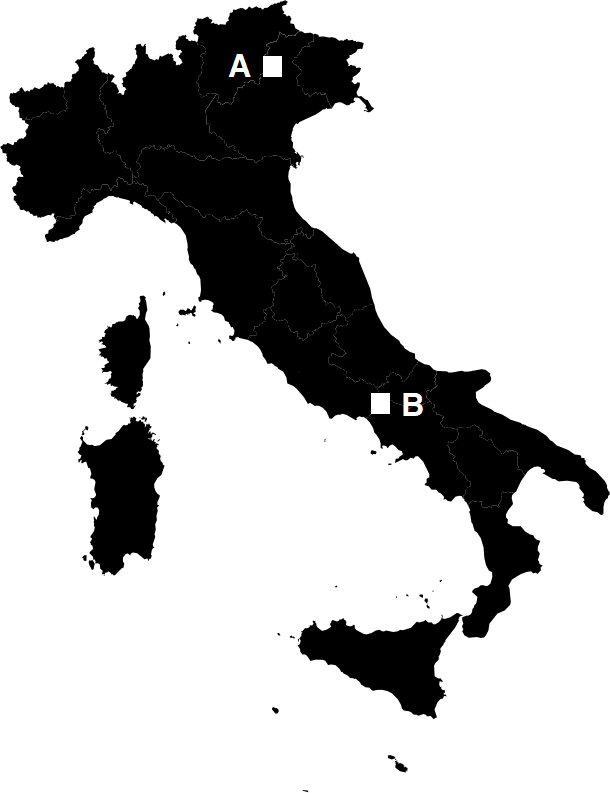}
	\caption{Location of the two considered hydropower plants in Italy.  
	}
	\label{fig:HPP_Plant_Soverzene_Presenzano_location}
\end{figure}

Throughout the paper, we will refer to two hydroelectric power plants, called plant A and plant B, which have an installed power of 215 MW and 1000 MW respectively. The plants are located in Italy as shown in Figure \ref{fig:HPP_Plant_Soverzene_Presenzano_location}, and both use Francis turbines. Plant A is of type reservoir, while plant B is of type pumped-storage. More details are provided in the following subsections.
\subsection{Hydropower plants details}
Plant A is located in Northern Italy and consists of four generation units moved by vertical axes Francis turbines, with a power of 60 MVA for each unit. The machinery room is located 500 meters inside the mountain. The plant is powered by two connected basins: the main basin, with a daily regulation purpose with a capacity of 5.9 $\times 10^6$ m$^3$, and a second basin, limited by a dam, with a seasonal regulation capability. The plant is part of a large hydraulic system and it has been operative since 1951. At full power, the plant employs a flow of 88 m$^3/$s, with a net head of 284 m; in nominal conditions (1 unit working 24/7, 3 units working 12/7) the main basin can be emptied in about 24 hours. The 2015 net production was of 594 GWh, serving both the energy and the service market thanks to the storage capability.

Plant B is representative of pumped-storage power plants; power is generated by releasing water from the upper reservoir down to the power plant which contains four reversible 250 MW Francis pump-turbine-generators. After power production, the water is sent to the lower reservoir. The upper reservoir, formed by an embankment dam, is located at an elevation of 643 meters in the Province of Isernia. Both the upper and lower reservoirs have an active storage capacity of 60 $\times 10^6$  m$^3$. The difference in elevation leads to a hydraulic head of 495 meters. The plant has been in operation since 1994, and its net production in 2015 was 60 GWh. This plant is strategical for its pumping storage capability and sells mainly to the services market.
\subsection{Dataset}
The dataset of plant A consists of 630  analog signals with a sampling time of 1 minute. The dataset of plant B consists of 60 analog signals, since a smaller number of sensors is installed in this system. The signals are collected from several plant-components, for instance, the water intake, penstocks, turbines, generators, and HV transformers. 
The acquisition system has been in service since the 1$^{st}$ of May 2017. In this work, we used data from the 1$^{st}$ of May 2017 to the 31$^{st}$ of March 2018, as the initial training set. Then, the model was tested online from the 1$^{st}$ of April 2018 until the end of July 2019. During the online testing phase, we  retrained the model every two months to include the most recent data.

As usual in this kind of applications, before starting the training phase, an accurate pre-processing was required to improve the quality of measured data. In particular:
\begin{enumerate}
\item 
several measured signals had a large number of not regular data, such as missing, or ``frozen'' samples (i.e., where the signal measured by the sensor does not change in time), values out of physical or operative limits, spikes, and statistical outliers in general;
\item 
the training dataset did not contain information about historical anomalies occurred in the plants; similarly, Operation and Maintenance (O\&M) logs were not available.
\end{enumerate}
Accordingly, we first implemented a classic procedure of data cleaning (see for instance \cite{Data_Cleaning_Book}). This procedure has two advantages itself: first of all, it allows any data-based condition monitoring methodology to be tuned upon nominal data corresponding to a correct functioning of the plant; in addition, the plants' operators were informed of those signals whose percentage of regular samples was below a given threshold, so that they could evaluate whether it could be possible to mitigate the noise with which data were recorded, or whether the sensor was actually broken.

The second problem was mainly due to the fact that in many operating plants, the O\&M logs are often not integrated with the historical databases. Thus, it may be subtle to distinguish nominal and faulty behaviours occurred in the past. In this case, we found it useful to iteratively merge analytic results with feedbacks from on-site operators to reconstruct correct sequences of nominal data. In particular, as we shall see in greater detail in the next section, the output of our proposed procedure is a newly proposed KPI, which monitors the functioning of the hydropower plant. In particular, a warning is triggered when the KPI drops below a threshold and an automatic notification is sent to the operator. The operator, guided by the provided warning, checks and possibly confirms the nature of the detected anomaly. Then, the sequence of data during the fault is eventually removed from the log, so that it will not be included in any historical dataset and it will not be used in the future retraining stages.

\section{Methodologies}
\label{Methodologies}
The proposed approach consists in training a self-organizing map (SOM) neural network in order to build a model of the nominal behaviour of the system, using a historical dataset comprehending nominal state observations. The new state observations are then classified as ``in control'' or ``out-of-control'' after comparing their distortion measure to the average distortion measure of the nominal states used during the training, as it is now illustrated in more detail.

\subsection{Self-Organizing Map neural network based Key Performance Indicator}
Self-organizing maps (SOMs) are popular artificial neural network algorithms, belonging to the unsupervised learning category, which have been frequently used in a wide range of applications \cite{Kohonen1990}-\cite{Tuc_2010}. Given a high-dimensional input dataset, the SOM algorithm produces a topographic mapping of the data into a lower-dimensional output set. 

The SOM output space consists of a fixed and ordered bi-dimensional grid of cells, identified by an index in the range $1,\dots,D$,  where a distance metric $d(c,i)$ between any two cells of index $c$ and $i$ is defined \cite{Kohonen1990}. Each cell of index $i$ is associated with a model vector $\mathbf{m}_i\in \mathbb{R}^{1 \times n}$ that lies in the same high-dimensional space of the input patterns $\mathbf{r}\in \Delta$, where the matrix $\Delta\in \mathbb{R}^{N \times n}$ represents the training dataset to be analyzed, containing $N$ observations of row vectors $\mathbf{r}\in \mathbb{R}^{1 \times n}$. After the training,  the distribution of the model vectors resembles the distribution of the input data, with the additional feature of preserving the grid topology: model vectors that correspond to neighbouring cells shall be neighbours in the high-dimensional input space as well. When a new input sample $\mathbf{r}$ is  presented to the network, the SOM finds the best matching unit (BMU) $c$, whose model vector $\mathbf{m}_c$ has the minimum Euclidean distance from $\mathbf{r}$:

\[c = \mathbf{arg min}_{i} \{\|   \mathbf{r} -  \mathbf{m}_i\|\}.\]

It is known that the goal of the SOM training algorithm is to minimize the following distortion measure:
\begin{equation}\label{DM_average}
DM_{\Delta}=\frac{1}{N} \sum_{\mathbf{r} \in \Delta} \sum_{i=1}^{D} w_{ci} \|\mathbf{r}-\mathbf{m}_i\|,
\end{equation}
where the function
\begin{equation}
w_{ci}  = exp\left( \frac{-d(c,i)^2}{2 \sigma^2} \right),
\end{equation}
is the neighbourhood function,  $c$ is the BMU corresponding to input sample $\mathbf{r}$, and $\sigma$ is the neighbourhood width. 
The distortion measure indicates the capacity of the trained SOM to fit the data maintaining the bi-dimensional topology of the output grid.
The distortion measure relative to a single input pattern $\mathbf{r}$  is computed as:
\begin{equation}\label{DM_single}
DM(\mathbf{r})= \sum_{i=1}^{D} w_{ci} \|\mathbf{r}-\mathbf{m}_i\|,
\end{equation}
from which it follows that $DM_{\Delta}$, as defined in (\ref{DM_average}), is the average of the distortion measures of all the patterns in the training data $\mathbf{r}\in \Delta$.

In order to  assess the condition of newly observed state patterns $\mathbf{r}$  to be monitored, we introduce the following KPI:
\begin{equation}
KPI(\mathbf{r})=\frac{1}{1+\|1- \frac{DM(\mathbf{r} )}{DM_{\Delta}}  \|}.
\end{equation}

Roughly speaking, the rationale behind the previous KPI definition is as follows: if the acquired state  $\mathbf{r}$ corresponds to a normal behaviour, its distortion measure $DM(\mathbf{r}) $ should be similar to the average distortion measure of the nominal training set $DM_{\Delta}$ (which consists of non-faulty states), and the ratio $ \frac{DM(\mathbf{r}) }{DM_{\Delta}} $ should be close to one, which in turn gives a $KPI( \mathbf{r})$ value to be close to one as well. On the other hand, if the acquired state $\mathbf{r}$ actually corresponds to an anomalous behaviour, $DM(\mathbf{r}) $ should substantially differ from $DM_{\Delta}$, leading to values of $KPI( \mathbf{r})$ considerably smaller than one. In this way, values of the KPI near to one indicate a nominal functioning, while smaller values indicate that the plant is going out of control.

A critical aspect is the choice of the threshold to discriminate a correct and a faulty functioning: for this purpose, we compute the average value $\mu_{KPI}$ and the variance $\sigma^2_{KPI}$ of the filtered KPI values of all the points in the training set $\Delta$, and we define the threshold as a lower control limit (LCL) as follows:

\begin{equation}
LCL_{kpi}= \mu_{kpi}-3\sigma_{kpi}.
\end{equation}

If the measured data are noisy, the proposed KPI may present a noisy nature as well. For this purpose, in our work we filtered the KPI using an exponentially weighted average filter over the last 12 hours of consecutive KPI values.

\subsection{The contribution of individual variables to the SOM-based KPI}

When the SOM-based KPI deviates from its nominal pattern, it is desired to identify the individual variables that most contribute to the KPI variation. This allows the operator not only to identify a possible malfunctioning in the hydropower plant, but also the specific cause, or location, of such a malfunctioning. For this purpose, we first calculate an average contribution to DM of individual variables using the data in the nominal dataset $\Delta$. We then compare the contribution of individual variables of newly acquired patterns to the average contribution of nominal training patterns.

For each pattern $\mathbf{r} \in \Delta$, we calculate the following average distance vector $\mathbf{d(r)}\in \mathbb{R}^{1 \times n}$ :

\begin{equation}
\mathbf{d(r)} =\frac{1}{D} \sum_{i=1}^{D} w_{ci} ( \mathbf{r}-\mathbf{m}_i ), \forall \mathbf{r} \in \Delta.
\end{equation}

Then we calculate the vector of squared components of $\mathbf{d(r)}$   normalized to have norm 1, named $\mathbf{d_n(r)} \in \mathbb{R}^{1 \times n}$, as

\begin{equation}\label{vet_dist}
\mathbf{d_n(r)} = \frac{\mathbf{d(r)}\circ \mathbf{d(r)} }{\| \mathbf{d(r)}\|^2}, \forall \mathbf{r} \in \Delta,
\end{equation}
where the symbol $\circ$ denotes the Hadamard (element-wise) product. Finally, we compute the average vector of the normalized squared distance components for all patterns in $\Delta$:

\begin{equation}
\mathbf{d_{n_{\Delta}}} =\frac{1}{N} \sum_{r\in\Delta} \mathbf{d_n(r)} .
\end{equation}

When a new pattern $\mathbf{r}$ is acquired during the monitoring phase, we calculate the following Hadamard ratio:

\begin{equation}
\mathbf{d_n(r)}\div \mathbf{d_{n_{\Delta}}} = [ cr_1 cr_2 \dots cr_n]
\end{equation}

where the contribution ratios $cr_i$ , $i=1\dots n$ represent how individual variables of the new pattern influence the DM compared to their influence in non-faulty conditions. If the new pattern actually corresponds to a non-faulty condition, $cr_i$ takes values close to one. If the new pattern deviates from the nominal behaviour, some of the  $cr_i$ exceed the unitary value. An empirical threshold 1.3 was selected, as a trade-off between false positives and true positives.

\subsection{Hotelling multivariate control chart}
As a term of comparison for our SOM-based KPI indicator, we consider the Hotelling multivariate control chart \cite{Hotelling1947}. While very few works may be found for our hydropower plant application of interest, Hotelling charts are quite popular for multivariable process control problems in general, and we take it as a benchmark procedure for comparison.

The Hotelling control chart performs a projection of the multivariate data to a scalar parameter denoted as $t^2$ statistics,
which is defined as the square of the Mahalanobis distance \cite{Maesschalck2000} between the
observed pattern and the vector containing the mean values of
the variables in nominal conditions. The Hotelling $t^2$ statistics is able to capture the changes in
multivariate data, revealing the deviations from the nominal
behaviour, and for these reasons the
Hotelling control chart is widely used for early detection of
incipient faults,  see for instance\cite{Aparisi2009}. 
The construction of the control chart includes two phases: in the
first phase, historical data are analyzed and the control limits
are computed; phase two corresponds to the monitoring of the newly acquired state patterns.
\subsubsection{Phase one}
 Let the nominal historic dataset be represented by the matrix ${{\Delta}} \in {\mathbb{R} ^{N \times n}}$,  containing N observations of row  vectors $ {{\bf{r}}}\in {\mathbb{R} ^{1 \times n}}$. The sample mean vector ${{\boldsymbol{\mu }}}\in {\mathbb{R} ^{1 \times n}}$ of the data  is defined as:
\begin{equation}
\label{mean_vector_phase_one}
{{{\boldsymbol{\mu }}}} = \frac{1}{N}\sum\limits_{\mathbf{r} \in \Delta} {{{\bf{r}}}} .
\end{equation}

The covariance matrix is defined by means of the zero-mean data matrix ${{\Delta}_{0}} \in {\mathbb{R} ^{N \times n}}$:
\begin{equation}
{{\Delta}_{0}} = \Delta- \bf{1} \cdot \boldsymbol{\mu},
\end{equation}
where ${\bf{1}} \in {\mathbb{R} ^{N \times 1}}$ represents a column vector with all entries equal to one.
Then, the covariance matrix ${{\bf{C}}} \in {\mathbb{R} ^{n \times n}}$ of the data is defined as:
\begin{equation}
{{\bf{C}}} = \frac{1}{{N - 1}}{\Delta}_{0}^T{{\Delta}_{0}},
\end{equation}
where  ${\left( {} \right)^T}$ denotes vector transpose  operation. The multivariate statistics  ${{\boldsymbol{\mu }}}$  and $\bf{C}$ represent the multivariate distribution of nominal observations, and we assume that  $\bf{C}$ is full rank.  The scalar ${t^2}$ statistics is defined as a function of a single state pattern ${\bf{r}}\in \mathbb{R} ^{1 \times n}$:
\begin{equation}
\label{t_2_def}
t^2(\bf{r}) = \left( {{{\bf{r}}} - {{\boldsymbol{\mu }}}} \right){\bf{C}}^{ - 1}{\left( {{{\bf{r}}} - {{\boldsymbol{\mu }}}} \right)^T}.\,
\end{equation}

The ${t^2}$ statistics is small when  pattern vector ${\bf{r}}$ represents nominal states, while it increases when the pattern vector ${\bf{r}}$  deviates from the nominal behaviour. In order to define the control limits $UCL$ and $LCL$ of the control chart, in this first phase we calculate the mean value $\mu_{t^2}$ and standard deviation $\sigma_{t^2}$ of the $t^2$ values obtained with all the observations of the historical dataset ${\bf{r}}\in{\Delta}$.  Then we define the safety thresholds as:
\begin{equation}
\label{Safety_Thresholds}
\left\{
\begin{array}{lll}
UCL_{t^2} & = & \mu_{t^2} + 3 \sigma_{t^2}\\
& & \\
LCL_{t^2} & = & max(\mu_{t^2} - 3 \sigma_{t^2}, 0)\\
\end{array}\right..
\end{equation}


\subsubsection{Phase two}
In the second phase, new observation vectors are measured, and the corresponding $t^2$  values are calculated as in Equation (\ref{t_2_def}). The Hotelling control chart may be seen as a monitoring tool that plots the $t^2$  values as consecutive points in time and compares them against the control limits. The process is considered to be ``out of control'',  when the $t^2$ values continuously exceed the control limits.

\section{Results}
\label{Results}
After a prototyping phase, the condition monitoring system has been operating since April 2018 on several components of the plants described in Section \ref{case_study}, with a total of more than 600 input signals. As an example, some of the most critical components are shown in Table \ref{tab:20190208_analyzed_components}. As can be seen from the last column of the table, there is usually a large redundancy of sensors measuring the same, or closely related, signals. The components listed in Table \ref{tab:20190208_analyzed_components} are among the most relevant components for the plant operators, as it is known that their malfunctioning may, in some cases, lead to major failures or plant emergency stops.

\begin{table}[ht!]
	\centering
	\begin{tabular}{l|l|c}
		\hline 
		 \textbf{Component name} & \textbf{Measured Signals} & \textbf{Number of sensors} \\
		\hline 
		 Generation Units & Vibrations & 34 \\
				\hline
		 HV Transformer & Temperatures & 27 \\
				 &  Gasses levels &  \\
				\hline
		 Turbine & Pressures & 27  \\
				 & Flows &  \\
				 & Temperatures &  \\
				\hline
		 Oleo-dynamic system & Pressures & 20 \\
				  & Temperatures &  \\
				\hline
		 Supports & Temperatures & 54  \\
				\hline
		 Alternator & Temperatures & 43  \\
		\hline
	\end{tabular}
	\vspace{0.05cm}
	\caption{List of main components analyzed for the hydro plants.}
	\label{tab:20190208_analyzed_components}
\end{table}


Since April 2018, our system detected more than $20$ anomalous situations, the full list of whose occurrences is given in Table \ref{occurrences}, with different degrees of severity, defined with plant operators, ranging from ``no action needed'' to ``severe malfunction'', leading to plant emergency stop.

It is worth noting that these events were not reported by the standard condition-monitoring systems operative in the plants, and in some cases would not have been well identified by the multivariate Hotelling control chart either, thus emphasizing the importance of the more sophisticated KPI introduced in this paper. In addition, we also see how the ability to identify non-nominal operations improves over time, as new information is acquired. 
\begin{table*}[htbp]
  \centering
     \begin{tabular}{|c|c|c|l|l|l|}
		\hline
		\rowcolor[rgb]{ .949,  .949,  .949} \textbf{Plant Code} & \textbf{Unit ID} & \textbf{Warning Date} & \textbf{Title} & \textbf{Severity Level} & \textbf{Feedback} \\
		\hline
    \rowcolor[rgb]{ .886,  .937,  .855} A     & 2     & 01/25/2018 & Efficiency Parameters & Low   & Weather anomalies effects  \\
		\hline
		\rowcolor[rgb]{ .886,  .937,  .855} A     & 1     & 07/02/2019 & Supports Temperature & Low   & Under investigation \\ 
		\hline
    \rowcolor[rgb]{ .886,  .937,  .855} A     & 3     & 03/03/2019 & Francis Turbine & Low   & Anomaly related to ongoing maintenance activities\\
		\hline
    \rowcolor[rgb]{ .988,  .894,  .839} A     & 1     & 06/11/2018 & Efficiency Parameters & Medium-Low & Not relevant anomaly \\
		\hline
    \rowcolor[rgb]{ .988,  .894,  .839} A     & 1     & 06/29/2018 & Francis Turbine & Medium-Low & Weather anomalies effects on turbine's components temperature \\
		\hline
    \rowcolor[rgb]{ .988,  .894,  .839} A     & 4     & 06/30/2018 & Efficiency Parameters & Medium-Low & Not relevant anomaly \\
		\hline
    \rowcolor[rgb]{ .988,  .894,  .839} A     & 2     & 08/07/2018 & Francis Turbine & Medium-Low & Weather anomalies effects on turbine's components temperature \\
		\hline
    \rowcolor[rgb]{ .988,  .894,  .839} A     & 2     & 09/14/2018 & Generator Vibrations & Medium-Low & Not relevant anomaly \\
		\hline
    \rowcolor[rgb]{ .988,  .894,  .839} A     & 3     & 01/10/2019 & Generator Temperature & Medium-Low & Under investigation \\
		\hline
    \rowcolor[rgb]{ .988,  .894,  .839} A     & 3     & 01/16/2019 & Transformer Temperature & Medium-Low & Weather anomalies effects on transformer temperature \\
		\hline
    \rowcolor[rgb]{ .988,  .894,  .839} A     & 2     & 03/03/2019 & Generator Temperature & Medium-Low & Anomaly related to ongoing maintenance activities\\
		\hline
    \rowcolor[rgb]{ .988,  .894,  .839} A     & 3     & 06/17/2019 & Generator Temperature & Medium-Low & Under investigation \\
		\hline
    \rowcolor[rgb]{ .988,  .894,  .839} A     & 4     & 06/21/2019 & Generator Temperature & Medium-Low & Weather anomalies effects on generator temperature \\	
		\hline
    \rowcolor[rgb]{ .886,  .937,  .855} A     & 1     & 04/22/2018 & HV transformer gasses & Medium-High & Monitoring system anomaly \\    
		\hline
    \rowcolor[rgb]{ .886,  .937,  .855} B     & 2     & 06/27/2018 & Generator Vibrations & Medium-High & Data Quality issue \\
		\hline
    \rowcolor[rgb]{ .886,  .937,  .855} A     & 1     & 10/29/2018 & Generator Vibrations & Medium-High & Data Quality issue \\
		\hline
    \rowcolor[rgb]{ .886,  .937,  .855} A     & 3     & 11/20/2018 & Generator Vibrations & Medium-High & Under investigation \\
		\hline
    \rowcolor[rgb]{ .886,  .937,  .855} A     & 1     & 03/24/2019 & Francis Turbine & Medium-High & Under investigation \\
		\hline
    \rowcolor[rgb]{ .886,  .937,  .855} A     & 2     & 04/07/2019 & Oleo-Dynamic System & Medium-High & Under investigation \\
		\hline
    \rowcolor[rgb]{ .988,  .894,  .839} B     & 2     & 10/01/2018 & Generator Temperature & High  & Sensor anomaly on block channel signal\\
		\hline
    \rowcolor[rgb]{ .988,  .894,  .839} B     & 2     & 11/01/2018 & Generator Temperature & High  & Sensor anomaly on block channel signal\\
		\hline
    \rowcolor[rgb]{ .988,  .894,  .839} A     & 2     & 03/01/2019 & HV transformer gasses & High  & Under investigation \\
		\hline
    \end{tabular}%
		\caption{List of anomalous behaviours that have been noticed during 16 months of test on the two hydropower plants. Lines 14 and 20 correspond to the two faults that have been illustrated in this paper.}
  \label{occurrences}%
\end{table*}%
 We now describe more in detail two different anomalies belonging to the two different plants, as summarized in Table \ref{tab:20190131_SOM_results}.
%
\begin{table}[ht!]
	\centering
	\begin{tabular}{c|c|c|l}
		\hline
		\textbf{Case Study} &\textbf {Plant}  & \textbf{Warning name}     &\textbf{ Warning date} \\
				\hline 
	 1	& B       & Generator Temperature  & 10/01/2018  \\
				\hline
	2	& A       & HV transformer gasses    & 04/22/2018 \\

		\hline
	\end{tabular}%
	\caption{Two failures reported by the predictive system and discussed in detail.} 
	\label{tab:20190131_SOM_results}%
\end{table}%
%


\subsection{Case Study 1: plant B - generator temperature signals}

\begin{figure}[!ht]
	\centering
	\includegraphics[trim=3cm 0cm 2cm 0cm,width=1\linewidth]
	{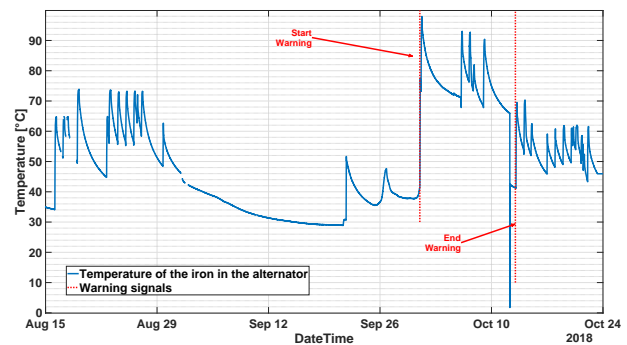}
	\caption{Measurement of the temperature sensor in the alternator of plant B. Anomalous values were detected by our system (start warning), timely actions were taken and the correct functioning was restored (end warning)  }
	\label{fig:Temperatura_1_Ferro_Alternatore_2_warning_id1_zoom.}
\end{figure}

In October 2018, an anomalous behavior was reported by our system, which indicated an anomaly regarding a sensor measuring the iron temperature of the alternator. It was then observed that the temperature values of such a sensor were higher than usual, as shown in Figure \ref{fig:Temperatura_1_Ferro_Alternatore_2_warning_id1_zoom.}, and also higher than the measurements taken by other similar sensors. However the temperature values did not yet exceed the warning threshold of the condition-monitoring systems already operative in the plant.

\begin{figure*}[!ht]
	\centering
	\includegraphics[trim=3cm 0cm 2cm 0cm,width=\linewidth]{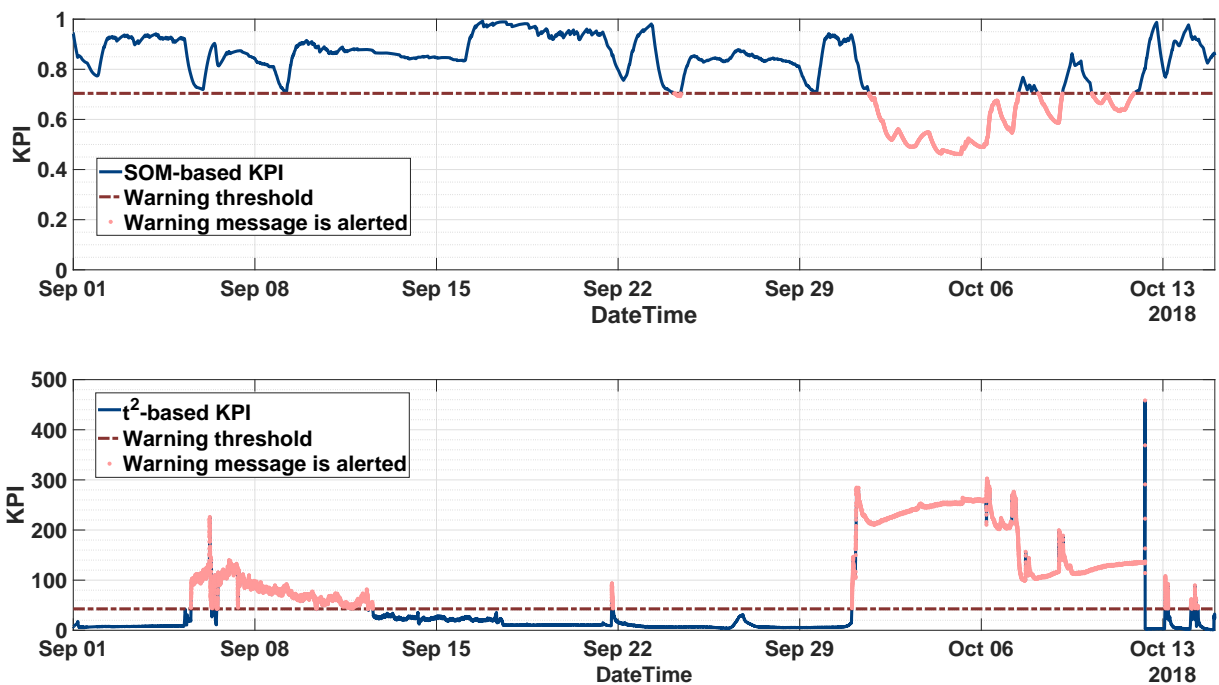}
	\caption{Plant A: KPI as a function of time. SOM-based results (top) are compared with $t^2$-based results (bottom).}
	\label{fig:KPI_temp_IDRT_subplot_SOM_T2}
\end{figure*}

From an inspection of the sensor measurements it was possible to establish that the exact day when this anomaly started occurring, the proposed SOM based KPI sharply notified a warning, as shown in Figure \ref{fig:KPI_temp_IDRT_subplot_SOM_T2};  for this case, the time-extension of the training dataset  was nine months, from 1 January 2018 to 1 September 2018. After receiving the warning alert, the plant operators checked the sensor and confirmed the event as a relevant anomaly. In particular, they acknowledged that this was a serious problem, since a further degradation of the measurement could have eventually led to the stop of the generation unit. For this reason, timely actions were taken: operators restored the nominal and correct behavior of the sensor starting on the 12$^{th}$ of October 2018.  The saved costs related to the prevented stopping of the generation unit were estimated in the range between 25 k\euro\ and 100 k\euro. 

While also the Hotelling control chart noticed an anomalous behaviour of the sensor in the same time frame, still it would have given rise to several false positives in the past, as shown in Figure \ref{fig:KPI_temp_IDRT_subplot_SOM_T2}.


%


\subsection{Case study 2: Plant A - HV Transformer anomaly}

\begin{figure*}[!ht]
	\centering
	\includegraphics[trim=4cm 0cm 2cm 0cm,width=\linewidth]
	{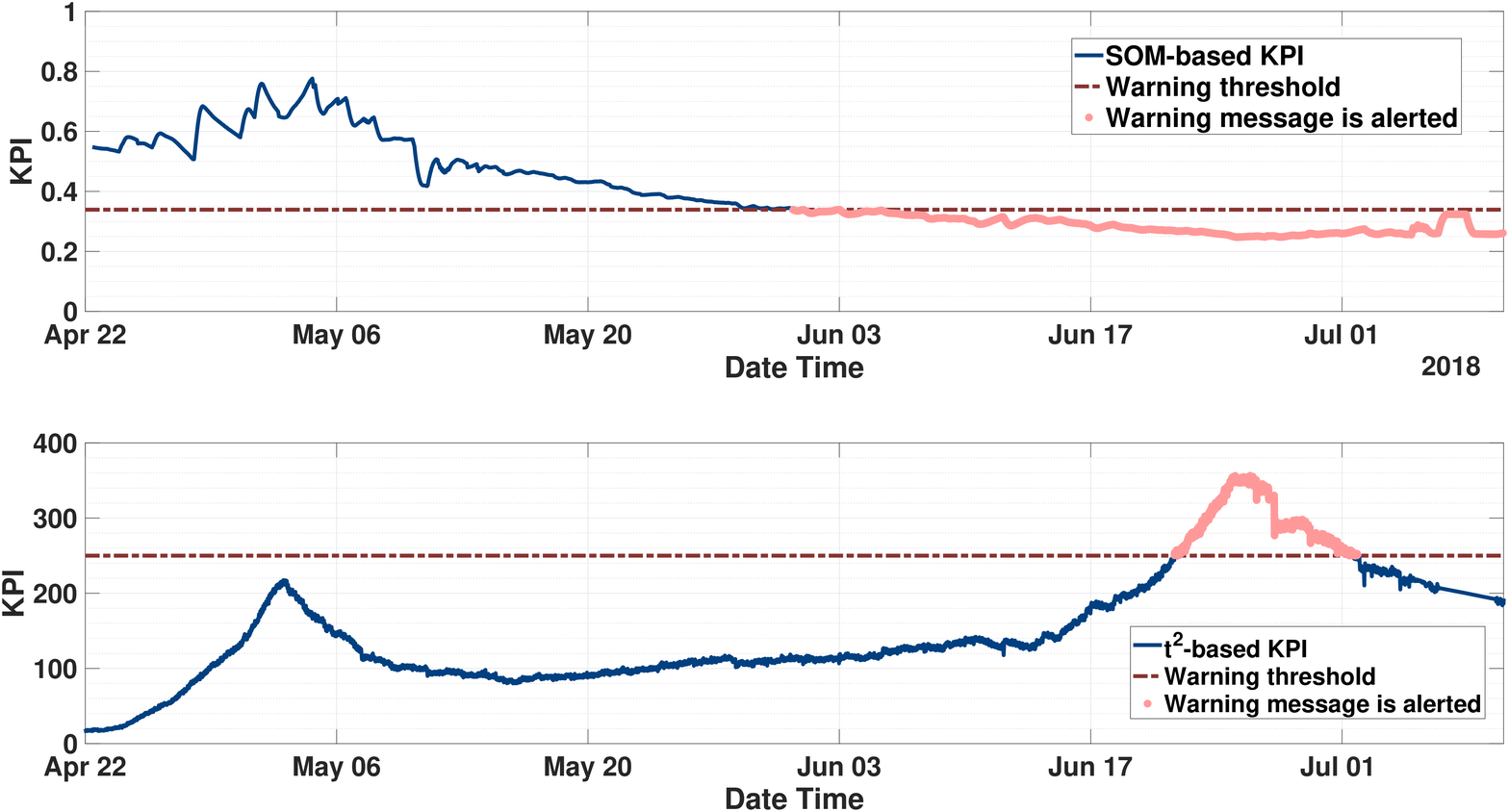}
	\caption{Plant B: KPI as a function of time. SOM-based results (top) are compared with $t^2$-based results (bottom).}
	\label{fig:KPI_hydran_IDCA_subplot_SOM_T2}
\end{figure*}

The SOM-based KPI detected  an anomaly on the HV transformer of one of the Generation Units at the beginning of June 2018 as shown in fig. \ref{fig:KPI_hydran_IDCA_subplot_SOM_T2}. After inspection of the operators, they informed us that similar anomalous situations had occurred in the past as well, but had not been tagged as faulty behaviours.  As soon as the time occurrences of the similar past anomalies had been notified by a plant operator, we proceeded to remove the corresponding signals from the training set. Then we recomputed our KPI based on the revised corrected historical dataset, and the KPI retrospectively found out that the ongoing faulty pattern had actually started one month earlier. This updated information was then validated by analyzing the output of an already installed gas monitoring system, that continuously monitors a composite value of various fault gases in ppM (Parts per Million) and tracks the oil-moisture. The gas monitoring system had been measuring increasing values, with respect to the historical ones, since the 22$^{nd}$ of April 2018, but no warning had been generated by that system. In this case, this was not however a critical fault, as the level of gasses in the transformer oil was not exceeding the maximum feasible limit. However, the warning triggered by our system was used to schedule maintenance actions that restored the nominal operating conditions. In addition, the feedback from the plant operators was very useful in order to tune the SOM-based monitoring system, and to increase its early detection capabilities. 

In this case, the Hotelling control chart realized of the anomalous behaviour only 20 days after our SOM-based KPI.

\section{Conclusion}
\label{Conclusion}
Driven by rapidly evolving enabling technologies, most notably Internet-of-Thing sensors and communication tools, together with more powerful artificial intelligent algorithms, condition monitoring, early diagnostics and predictive maintenance methodologies and tools are becoming some of the most interesting areas of research in the power community. While some preliminary examples can be found in many fields, solar and wind plants being one of them, fewer applications can be found in the field of hydropower plants. In this context, this paper is one of the first examples, at least up to the knowledge of the authors, that results of a newly proposed KPI are validated over a 1-year running test field in two hydropower plants.\\
\newline
This paper provided very promising preliminary results, that encourage further research on this topic. In particular the proposed procedure can not be implemented in a fully unsupervised fashion yet, but some iterations with plant operators still take place when alarm signals are alerted. Also, the proposed condition monitoring strategy is a first step towards fully automatic predictive maintenance schemes, where faults are not just observed but are actually predicted ahead of time, possibly when they are only at an incipient stage. In the opinion of the authors this is a very promising area of research and there is a general interest towards the development of such predictive strategies.

\ifCLASSOPTIONcaptionsoff
  \newpage
\fi

\begin{IEEEbiography}[{\includegraphics[width=1in,height=1.25in,clip,keepaspectratio]{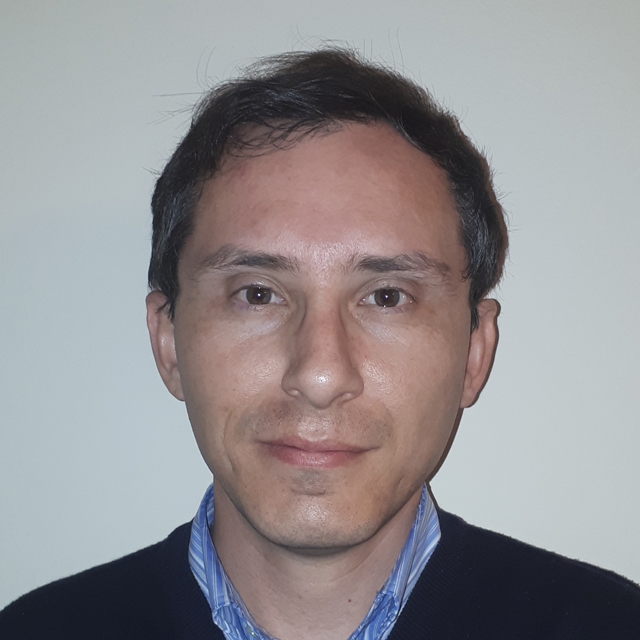}}]{Alessandro Betti}
	Dr. Alessandro Betti received the M.S. degree in Physics and the Ph.D. degree in Electronics Engineering from the University of Pisa, Italy, in 2007 and 2011, respectively. His main field of research was modeling of noise and transport in quasi-one dimensional devices. His work has been published in 10 papers in peer-reviewed journals in the field of solid state electronics and condensed matter physics and in 16 conference papers, presenting for 3 straight years his research at the top International Conference IEEE in electron devices, the International Electron Device Meeting in USA. In September 2015 he joined the company i-EM in Livorno, where he currently works as a Senior Data Scientist developing power generation forecasting, predictive maintenance and Deep Learning models, as well as solutions in the electrical mobility fields and managing a Data Science Team.
\end{IEEEbiography}

\begin{IEEEbiography}[{\includegraphics[width=1in,height=1.25in,clip,keepaspectratio]{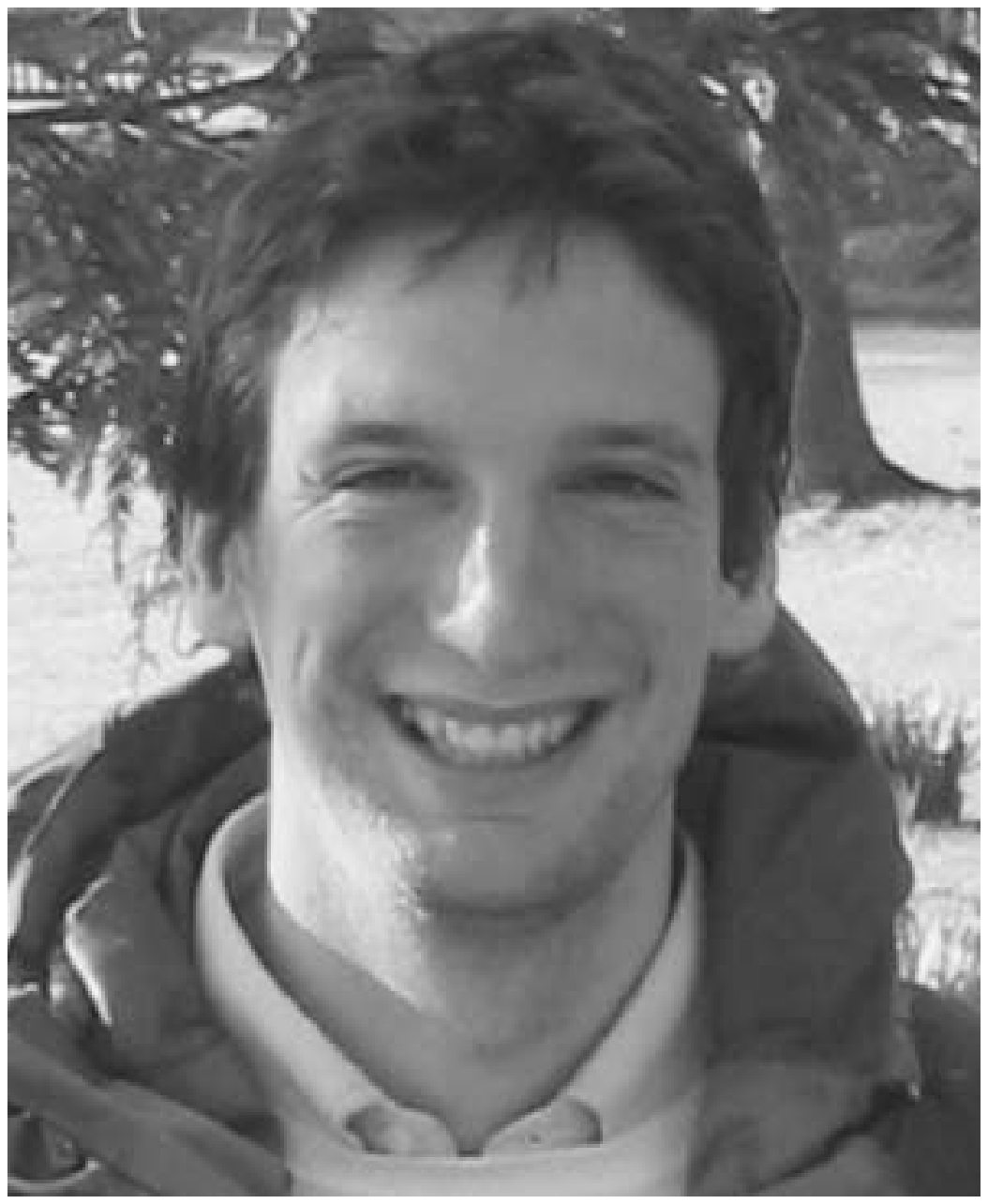}}]{Emanuele Crisostomi}(M'12-SM'16) received the B.Sc. degree in computer science engineering, the M.Sc. degree in automatic control, and the Ph.D. degree in automatics, robotics, and bioengineering from the University of Pisa, Pisa, Italy, in 2002, 2005, and 2009, respectively. He is currently an Associate Professor of electrical engineering in the Department of Energy, Systems, Territory and Constructions Engineering, University of Pisa. He has authored tens of publications in top refereed journals and he is a co-author of the recently published book on ``Electric and Plug-in Vehicle Networks: Optimization and Control'' (CRC Press, Taylor and Francis Group, 2017). His research interests include control and optimization of large scale systems, with applications to smart grids and green mobility networks. 
	
\end{IEEEbiography}

\begin{IEEEbiography}[{\includegraphics[width=1in,height=1.25in,clip,keepaspectratio]{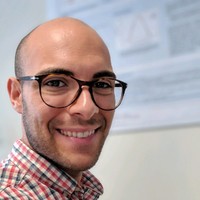}}]{Gianluca Paolinelli}

Gianluca Paolinelli received the bachelor and master degree in electrical engineering from the University of Pisa, Pisa, Italy, in 2014 and 2018 respectively. His research interests included big data analysis and computational intelligence applied in on-line monitoring and diagnostics.
Currently, he is an electrical software engineer focused in the development of electric and hybrid power-train controls for Pure Power Control S.r.l.  

\end{IEEEbiography}

\begin{IEEEbiography}[{\includegraphics[width=1in,height=1.25in,clip,keepaspectratio]{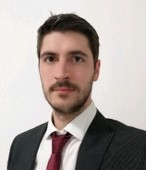}}]{Antonio Piazzi}
	Antonio Piazzi received the M.Sc. degree in electrical engineering from the University of Pisa, Pisa, Italy, 2013. Electrical engineer at i-EM since 2014, he gained his professional experience in the field of renewable energies. His research interests include machine learning and statistical data analysis, with main applications on modeling and monitoring the behaviour of renewable power plants. Currently, he is working on big data analysis applied on hydro power plants signals.		
\end{IEEEbiography}

\begin{IEEEbiography}[{\includegraphics[width=1in,height=1.25in,clip,keepaspectratio]{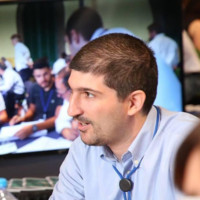}}]{Fabrizio Ruffini}
	Fabrizio Ruffini received the Ph.D. degree in Experimental Physics from University of Siena, Siena, Italy, in 2013. His research-activity is centered on data analysis, with particular interest in multidimensional statistical analysis. During his research activities, he was at the Fermi National Accelerator Laboratory (Fermilab), Chicago, USA, and at the European Organization for Nuclear Research (CERN), Geneva, Switzerland. Since 2013, he has been working at i-EM as data scientist focusing on applications in the renewable energy sector, atmospheric physics, and smart grids. Currently, he is senior data scientist with focus on international funding opportunities and dissemination activities.
\end{IEEEbiography}

\begin{IEEEbiography}[{\includegraphics[width=1in,height=1.25in,clip,keepaspectratio]{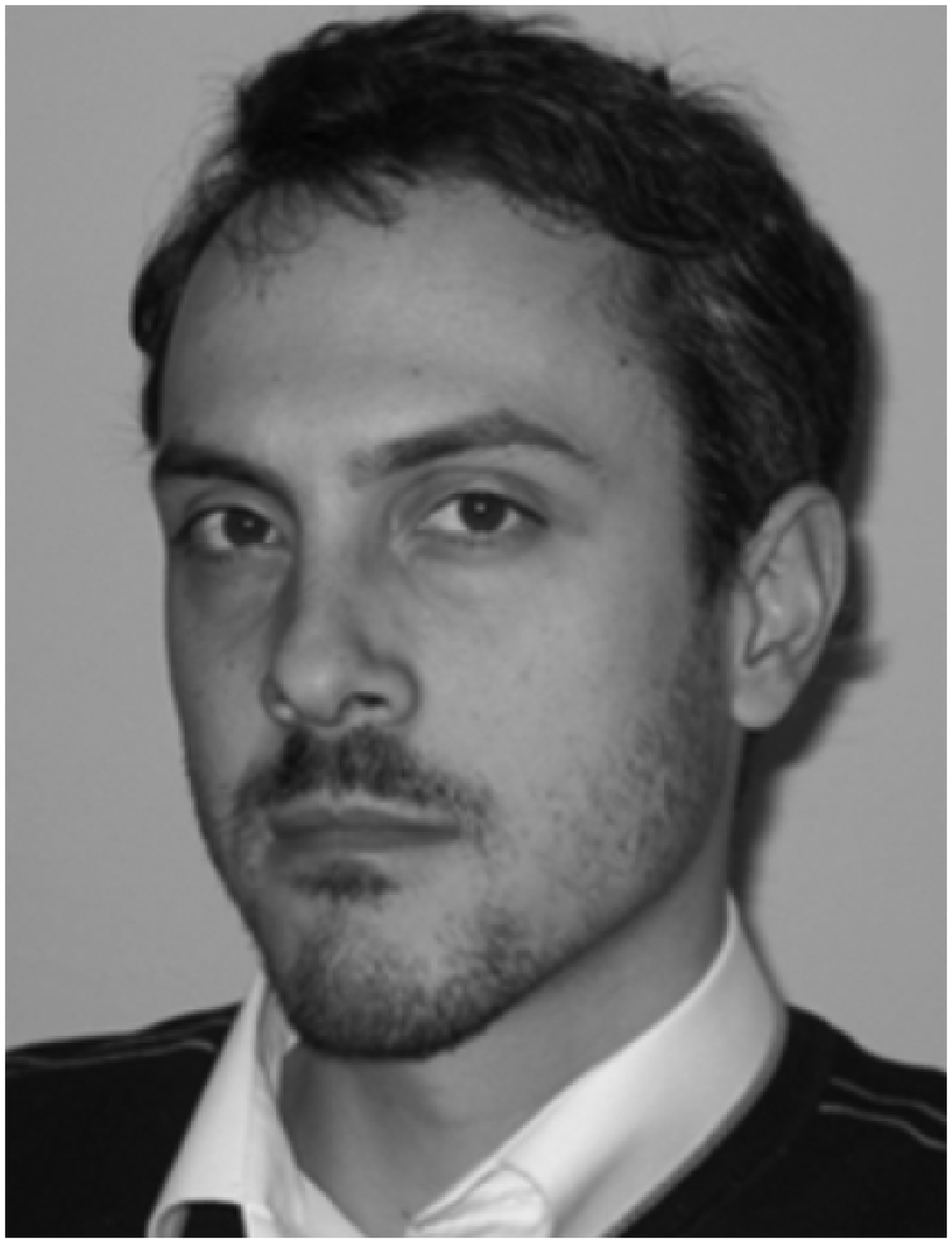}}]{Mauro Tucci} received the Ph.D.degree in applied electromagnetism from the University of Pisa, Pisa, Italy, in 2008. Currently, he is an Associate Professor with the Department of Energy, Systems, Territory and Constructions Engineering, University of Pisa. His research interests include computational intelligence
and big data analysis, with applications in electromagnetism, non destructive testing, powerline communications, and smart grids.
\end{IEEEbiography}




\end{document}